\documentclass[british,prd,floatfix,superscriptaddress,nofootinbib]{revtex4-2}
\usepackage[T1]{fontenc}
\usepackage[utf8]{inputenc}
\usepackage{color}
\definecolor{note_fontcolor}{rgb}{0.800781, 0.800781, 0.800781}
\usepackage{babel}
\usepackage{mathtools}
\usepackage{amsmath}
\usepackage{soul}
\usepackage{amssymb}
\usepackage{booktabs}
\usepackage{graphicx}
\PassOptionsToPackage{hyphens}{url}
\usepackage[unicode=true,pdfusetitle,
bookmarks=true,bookmarksnumbered=false,bookmarksopen=false,
breaklinks=false,pdfborder={0 0 0},pdfborderstyle={},backref=false,colorlinks=false]{hyperref}

\makeatletter

\usepackage{bm}
\usepackage{upgreek}
\usepackage{siunitx}
\usepackage{diagbox}
\usepackage{lipsum}
\usepackage{epigraph}
\usepackage{url}

\usepackage{tensind}
\tensordelimiter{?}
\usepackage{tensor}
\usepackage{pgfplots}
\pgfplotsset{compat=1.18}
\usetikzlibrary{pgfplots.groupplots}

\DeclareSymbolFont{vectors}{OML}{cmm}{b}{it}
\DeclareSymbolFont{tensors}{OT1}{cmss}{bx}{n}

\DeclareSymbolFontAlphabet{\mathvec} {vectors}
\DeclareSymbolFontAlphabet{\mathtens}{tensors}

\usepackage{cleveref}

\ifdefined\pdftexversion\else  
  \usepackage{fontspec}
\fi
\@ifpackageloaded{underscore}{}{\usepackage[strings]{underscore}}\makeatother

\makeatother

\begin{document}

\global\long\def\tudu#1#2#3#4{{?{#1}^{#2}{}_{#3}{}^{#4}?}}%

\global\long\def\tud#1#2#3{{?{#1}^{#2}{}_{#3}?}}%

\global\long\def\tudud#1#2#3#4#5{{?{#1}^{#2}{}_{#3}{}^{#4}{}_{#5}?}}%

\global\long\def\tdu#1#2#3{\tensor{#1}{_{#2}^{#3}}}%

\global\long\def\dd#1#2{\frac{\mathrm{d}#1}{\mathrm{d}#2}}%

\global\long\def\pd#1#2{\frac{\partial#1}{\partial#2}}%

\global\long\def\tens#1{\mathtens{#1}}%

\global\long\def\threevec#1{\mathvec{#1}}%

\global\long\def\d{\mathrm{d}}%

\global\long\def\e{\mathrm{e}}

\global\long\def\eps{\varepsilon}

\global\long\def\i{\mathrm{i}}

\global\long\def\ext{\tilde{\mathrm{d}}}%

\title{Stability and collisions of excited spherical boson stars:\\ glimpses of chains and rings}
\author{Marco Brito}
\email[]{marcobrito@ua.pt}
\affiliation{Departamento de Matemática da Universidade de Aveiro \\
    and Centre for Research and Development in Mathematics and Applications (CIDMA) \\
    Campus de Santiago, 3810-193 Aveiro, Portugal}
\author{Carlos Herdeiro}
\email[]{herdeiro@ua.pt}
\affiliation{Departamento de Matemática da Universidade de Aveiro \\
    and Centre for Research and Development in Mathematics and Applications (CIDMA) \\
    Campus de Santiago, 3810-193 Aveiro, Portugal}
\author{Eugen Radu}
\email[]{eugen.radu@ua.pt}
\affiliation{Departamento de Matemática da Universidade de Aveiro \\
and Centre for Research and Development in Mathematics and Applications (CIDMA) \\
Campus de Santiago, 3810-193 Aveiro, Portugal}
\author{Nicolas Sanchis-Gual}
\email[]{nicolas.sanchis@uv.es}
\affiliation{Departamento de Astronomía y Astrofísica, Universitat de València, \\
    Dr.\ Moliner 50, 46100, Burjassot (Valencia), Spain}
\author{Miguel Zilhão}
\email[]{mzilhao@ua.pt}
\affiliation{Departamento de Matemática da Universidade de Aveiro \\
    and Centre for Research and Development in Mathematics and Applications (CIDMA) \\
    Campus de Santiago, 3810-193 Aveiro, Portugal}

\begin{abstract}
     Scalar, spherically symmetric, radially excited boson stars were previously shown to be stabilized, against spherical dynamics, by sufficiently strong self-interactions. Here, we further test their stability now in a full 3+1D evolution. We show that the stable stars in the former case become afflicted by a non-spherical  instability.
     Then, we perform head-on collisions of  both (stable) fundamental and (sufficiently long-lived) excited boson stars. Depending on the stars chosen,  either a black hole or a bosonic remnant are possible. In particular, collisions of excited stars result in a bosonic bound state which resembles a dynamical superposition of chains and rings, akin to the ones found as equilibrium solutions in~\cite{Liang:2025myf}. These evolutions emphasize a key difference concerning the dynamical robustness of fundamental \textit{vs} excited spherical boson stars, when generic (beyond spherical) dynamics is considered.
\end{abstract}
\maketitle

\section{Introduction} 

Bosonic stars are solutions to the Einstein-scalar or Einstein-Proca equations, representing a self-gravitating lump of
(massive) bosonic fields sustained by their own self-gravity. The bosonic field can be free or self-interacting~\cite{Schunck:2003kk,Liebling:2012fv}.
In some models, their dynamics have been proposed to match real gravitational wave signals \cite{CalderonBustillo:2022cja,CalderonBustillo:2020fyi}.
They have also been proposed as dark matter candidates
\cite{Lee:1995af,Suarez:2013iw,Eby:2015hsq,Chen:2020cef}. Additionally, bosonic stars can also be exotic compact objects, mimicking the phenomenology of black holes
\cite{Schunck:1998cdq,Mielke:2000mh,Berti:2006qt,Guzman:2009zz,Vincent:2015xta,Grould:2017rzz,Olivares:2018abq,Herdeiro:2021lwl,Rosa:2022tfv,Sengo:2024pwk,bezares2025exotic}.

Focusing, for concreteness, on spherically symmetric scalar boson stars,
one can find an infinitude of solutions, labelled by the number of radial nodes $n$, ranging from the fundamental state ($n=0$) to excited states ($n\geqslant 1$),
analogous to the $s$-orbitals of the hydrogen atom. This is only possible in the scalar case, since the
fundamental state for vector (Proca) stars is not spherically symmetric \cite{Herdeiro:2023wqf}. Moreover, recently it has been found that the excited spherical stars can bifurcate, in the space of equilibrium solutions, into families of chains and rings~\cite{Liang:2025myf}. So far, however, no dynamical connection between these different solutions has been found.

Further focusing on a scalar model with a quartic potential (see \cite{Colpi:1986ye,Bernal:2009zy,Hartmann:2013tca,Herdeiro:2017fhv,Alcubierre:2018ahf} for other models), it was shown in~\cite{Brito:2023fwr,Sanchis-Gual:2021phr} that scalar, spherical excited
stars, previously thought to be unstable \cite{Balakrishna:1997ej}
could be stabilized by an adequate strong enough self-interaction {similarly to the rotating case~\cite{Sanchis-Gual:2019ljs,siemonsen2021stability}}. The evolutions
performed, however, imposed spherical symmetry. {Moreover, it was shown \cite{Nambo:2024gvs} that these excited boson stars, in the Newtonian limit, are unstable with respect to non-spherical perturbations.} Thus, the question remains regarding their stability
with respect to generic perturbations (see~\cite{jaramillo2020dynamical}), {in the relativistic regime}. 
If such stars are stable -- or their instability timescale is large compared to other relevant dynamical timescales -- one may inquire about the end result of collisions between these objects. Namely, if it
would generate an excited bosonic remnant, a boson star in the fundamental
state, collapse to a black hole or a different outcome. To assess these dynamics is one of the goals of this work.

In particular, 
it was reported in \cite{Sanchis-Gual:2021phr} that excited configurations (for the $n=1$ case
only) could be formed by a gravitational cooling mechanism starting from an appropriate spherically symmetric dilute
Gaussian distribution. It is {therefore} interesting to study if these excited states can also be formed via
collisions, like their fundamental counterparts. As we shall see, our evolutions unveil a non-spherical instability of the excited boson stars and also show glimpses of an unexpected connection with the aforementioned chains and rings, for collisions of (sufficiently long-lived) excited spherical stars.

The paper is organized as follows: in \Cref{model} we describe the action and equations of motion
that generate these bosonic stars as equilibrium solutions, as well as the numerical framework used for the evolutions.
In \Cref{stability} we analyze single boson star evolutions to ascertain their stability with
respect to generic perturbations. In \Cref{collisions} we perform head-on collisions between excited boson stars, observing
collisions that result in a black hole and other collisions resulting in a bosonic remnant. We
close with a discussion and some final remarks.

\section{The model} \label{model}

\subsection{The action and equations of motion}
We consider a massive self-interacting scalar field coupled to Einstein's gravity given by the action
\begin{equation}
    S_{\mu,\lambda}[\mathtens{g},\Phi,\Phi^{*}] = \int_{{\cal M}}\d ^{4}x \sqrt{-g} \left[\frac{R}{16\pi }-
        \frac{1}{2}\left(\Phi_{,\alpha}^{*}\Phi^{,\alpha}+\mu^{2}|\Phi|^{2}+\frac{\lambda}{2}|\Phi|^{4}\right)\right],
\end{equation}
where $\lambda$ is the self-interaction coupling constant. In order to rescale the quantities by
$\mu$, we define the dimensionless self-coupling constant $\Lambda:= \lambda /(4\pi\mu^2)$. Throughout this article, we shall use the metric signature $(-,+,+,+)$ and
use geometrized units $c=G=1$.
The corresponding equations of motion are 
\begin{align}
    G_{\alpha\beta} =8\pi T_{\alpha\beta},  \qquad
    \Box\Phi  =\left(\mu^{2}+4\pi\mu^2\Lambda|\Phi|^{2}\right)\Phi,
\end{align}
where the stress-energy tensor is defined as
\[T_{\alpha\beta}=\Phi_{,(\alpha}^{*}\Phi_{,\beta)}-\frac{1}{2}g_{\alpha\beta}
    (\Phi_{,\rho}^{*}\Phi^{,\rho}+\mu^{2}|\Phi|^{2}+2\pi\mu^2\Lambda|\Phi|^{4}).\]
We set $\mu=1$ in most of the presentation below (i.e., quantities like the Arnowitt-Deser-Misner (ADM) mass and frequency are assumed to be scaled by $\mu$), but keep it in the figures, for clarity. $\mu$ is referred to as the mass parameter; more precisely, it is the inverse Compton wavelength. 

We can also write the Komar mass for asymptotically flat and stationary spacetimes as
\[M_K = \int \d x \d y \d z \sqrt{-g}\ \rho_K,\]
where $\rho_K\coloneqq T-2T^t_t$ is the Komar energy density, which allows us to compute the ADM mass of our initial data.

\subsection{Ansatz}
We are interested in spherically symmetric configurations, as in \cite{Brito:2023fwr},
which can be described by the following metric ansatz in isotropic coordinates
\begin{equation} 
\label{metric-isotropic}
    \d s^2=-\e^{2F_0 (r)}\d t^2 +\e ^{2F_1 (r)}[\d r^2 +r^2(\d \theta^2+\sin^2\theta\d\phi^2)] \ ,
\end{equation}
and the scalar field ansatz
\begin{equation}
    \Phi(x^\alpha)=\phi(r)\e^{-\i \omega t},
\end{equation}
where $\omega>0$ is the frequency of the field.


To find equilibrium configurations describing spherical boson stars, we solve the Einstein-Klein-Gordon system with the above ansatz, leading to second-order ordinary differential equations for the functions
$F_{0},F_{1}$ and $\phi$. These are solved numerically, imposing two boundary conditions for each function. The boundary conditions at the origin, to ensure
regularity, are given by
\[
    \partial_{r}F_{0,1}(0)=0,\quad\partial_{r}\phi(0)=0 ,
\]
whereas asymptotic flatness requires
\[
    F_{0,1}(r\to\infty)=\phi(r\to\infty)=0.
\]

\subsection{Solutions and evolution set-up}

\begin{figure}[htbp]
    \centering
    \input{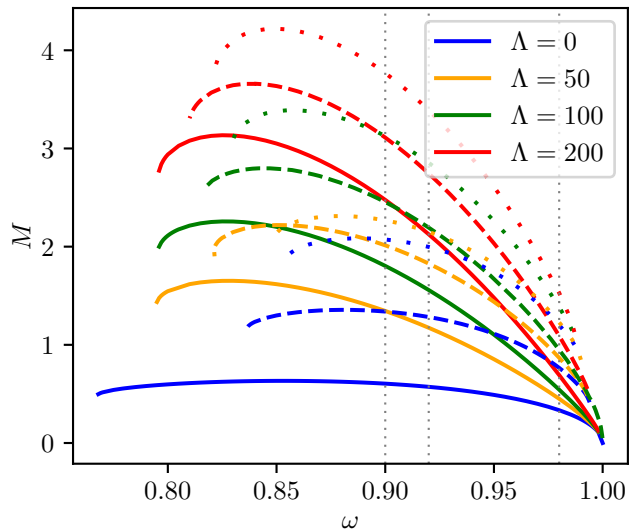}
    \caption{ADM mass $vs.$ the frequency for boson star solutions with four different values of $\Lambda$ and different values of the radial excitation number $n$. Solid curves correspond to fundamental boson stars, dashed curves to $n=1$ and dotted curves to $n=2$. The vertical lines indicate the frequencies of the solutions we considered.}
    \label{M-vs-w}
\end{figure}

Solving the equations above will result in a set of infinite families of solutions, with each family labelled by a different number of radial nodes $n$.
For each family, the solutions only exist for values of the frequency
between $\omega_{\rm{min}} < \omega < 1$. The solutions are located on
curves -- often called \textit{solution curves} -- such as the ones given in \Cref{M-vs-w}, 
which relate the
ADM mass with the frequency
of the stars.

The system above was solved to obtain a continuous family of solutions of scalar boson stars for
each $\Lambda$, parameterized by $\omega$. This is done by finding the
solutions to the coupled non-linear differential equations for the functions $\mathcal{F} = (F_0, F_1; \phi)$ using 
a professional software
package which employs a finite difference method and the  Newton-Raphson algorithm
\cite{Schoenauer1989}.

An important property of boson star solutions is that the scalar field only becomes zero at infinity. In this sense they do not possess a surface. Nevertheless, it is customary to define an \textit{effective (areal) radius}, {$R_{99}$}, by determining it should contain, say, 99\% of the mass of the star. In the following, we shall use this definition for the radius of boson stars. Since we shall work with isotropic coordinates -- eq.~\eqref{metric-isotropic} -- the corresponding isotropic radius is denoted $r_{99}$. {We define the compactness of the star as $C_{99}=0.99M_\mathrm{ADM}/R_{99}$.} 
 \Cref{solutions} gives relevant physical quantities for the solutions analyzed in this article.
\begin{table}[htbp] 
    \centering
\begin{tabular}{cccccccccc}
\toprule 
Solution \# & I & II & III & IV & V & VI & VII & VIII & IX \\
\midrule
$[n,\omega,\Lambda]$ & $[0,0.90,20]$ & $[0,0.90,125]$ & $[1,0.92,100]$ & $[1,0.92,200]$ & $[1,0.92,250]$ & $[2,0.92,200]$ & $[0,0.98,0]$ & $[1,0.98,100]$ & $[2,0.98,100]$ \\
\midrule
$r_{99}$ & $12.664$ & $21.658$ & $30.423$ & $36.407$ & $38.901$ & $44.545$ & $26.942$ & $57.941$ & $79.477$ \\
\midrule 
$M_{\rm ADM}$ & $0.968$ & $1.659$ & $1.924$ & $2.745$ & $2.972$ & $3.388$ & $0.335$ & $0.952$ & $1.371$ \\
\midrule 
$C_{99}$ & $0.070$ & $0.084$ & $0.067$ & $0.070$ & $0.070$ & $0.069$ & {$0.012$} & {$0.016$} & {$0.017$} \\
\bottomrule
\end{tabular}
\caption{Physical quantities of the boson stars used in this article.}
\label{solutions}
\end{table}

For the numerical evolutions we cast the equations of motion in the BSSN (Baumgarte-Shapiro-Shibata-Nakamura) form~\cite{Baumgarte:1998te, Shibata:1995we} and evolve them using the \textsc{EinsteinToolkit} infrastructure~\cite{Loffler:2011ay,Zilhao:2013hia,EinsteinToolkit:2025-05}
with the Carpet library for mesh refinement capabilities~\cite{Schnetter:2003rb,CarpetCode:web}.
Our implementation follows the approach detailed in~\cite{Cunha:2017wao,Okawa:2014nda} and we use the 
LeanBSSNMoL and ScalarEvolve \textsc{Cactus} \emph{thorns} 
to evolve the metric and scalar field variables in time. These codes are publicly available~\cite{canuda-2023-7791842} and have been employed and described in previous works~\cite{Cunha:2017wao,Sanchis-Gual:2019ljs,Sanchis-Gual:2020mzb,DiGiovanni:2020frc,Ildefonso:2023qty}.

Regarding the evolutions of single stars, the evolutions
are performed in a Cartesian grid with $512\times512\times512$ cells and
four levels of grid refinement -- the
outermost one having a resolution of either $\d x=2$ or $\d x=8/3$
where the resolution
doubles
through each refinement level. The innermost refinement level boundaries
are adjusted so that the effective radius of the star is localized inside such
level, but no larger than needed in order to save computational resources. For the very same purpose, we impose a $\mathbb{Z}_2$ symmetry with respect to the $z=0$ plane, and for longer  (than $\simeq 3000$) evolutions we also impose a $\mathbb{Z}_2$ symmetry with respect to the $y=0$ plane. There is no difference between the results with or without the imposed symmetries.

Our initial data is constructed through a superposition of the aforementioned (static) spherically symmetric solutions, using a scheme of the form~\cite{Helfer:2018vtq,Helfer:2021brt} used in many recent works~\cite{bezares2022gravitational,ge2024hair,palloni2025eccentric}
\begin{equation}
    \gamma_{ij}(x^{k})=\gamma_{ij}^{A}(x^{k})+\gamma_{ij}^{B}(x^{k})-h_{ij}, \label{sup-eq} 
\end{equation}
where we make $h_{ij}=\gamma_{ij}^{B}(x_{A}^{k})$, and $x^k_A$ are the coordinates of the centre of star
A\@. This differs from the most straightforward superposition scheme, by having $h_{ij}=\gamma_{ij}^{B}(x_{A}^{k})$ instead of simply $h_{ij}=\delta_{ij}$. This construction avoids spurious oscillations in the remnant object, which were due to the volume element at the centre of the stars differing from their equilibrium values. The 
violation of the Hamiltonian constraint is also improved, especially at the centre of
both stars. This choice, however, has the disadvantage that asymptotic flatness is not preserved exactly, unlike in the case where $h_{ij}=\delta_{ij}$.

\begin{figure}[htbp]
    \centering
    \includegraphics[scale=0.25]{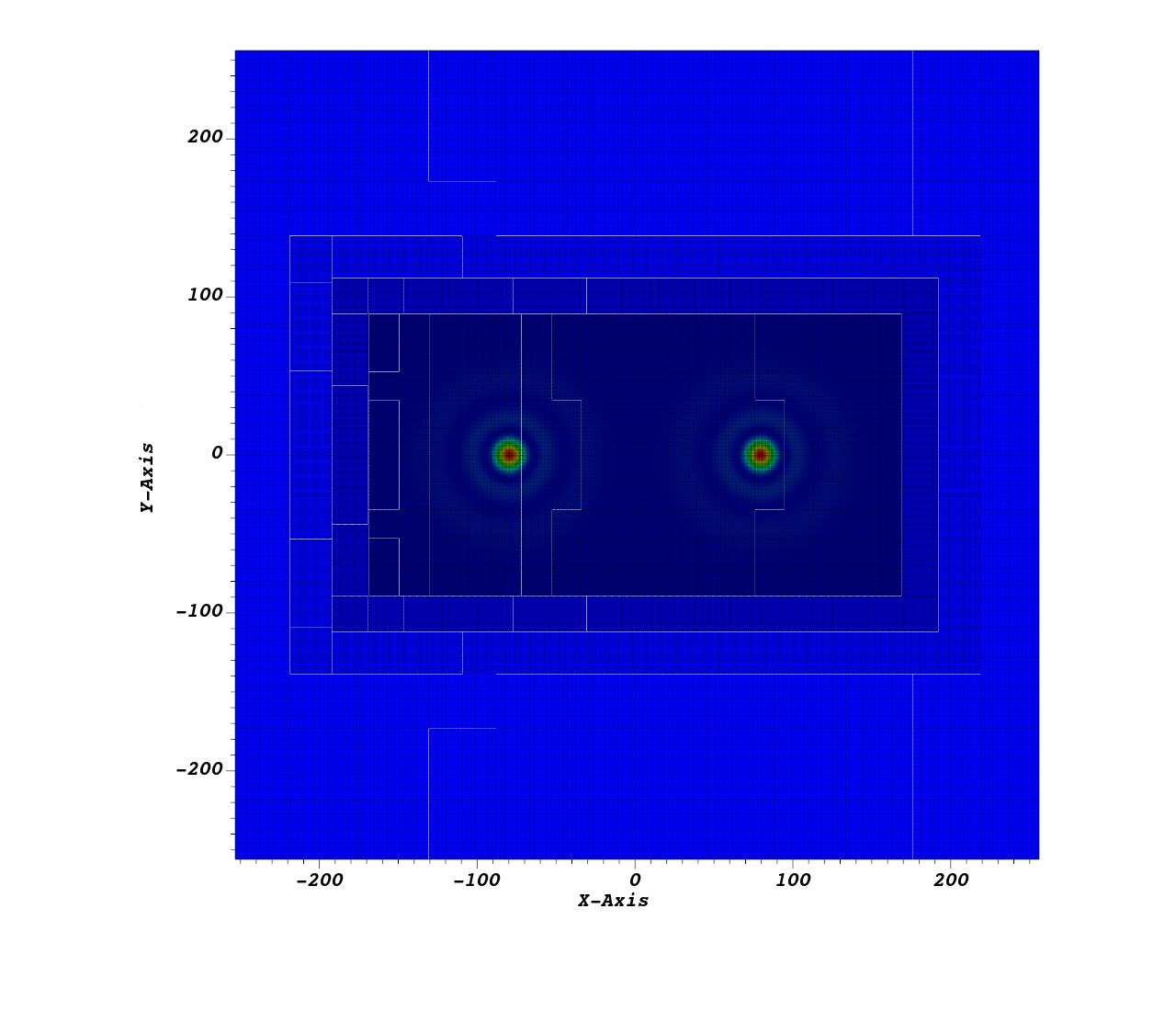}
    \caption{Grid structure, in units of $\mu$, for a head-on collision of stars with $n=2,\ \omega=0.98,\ \Lambda=100$ (solutions IX), and the energy density of the scalar field ($\mu^{-2}\rho_K$).}
    \label{fig:grid}
\end{figure}

No specific
perturbation was applied since numerical error suffices to break the
staticity of the models, in case instabilities are present. As for the evolution
of binaries in a head-on collision we used three numerical centres with one
located at the centre of the grid and the others with their centres at the centre
of each star, so that they overlap and make a single rectangle where the refinement levels match -- see \Cref{fig:grid}.
The latter numerical centres have four refinement levels while the one at
the centre may have more levels of refinement in order to properly resolve length scales in cases where a black hole remnant after the collision is obtained.

\section{Stability in a 3+1 setting} \label{stability}

In a previous work \cite{Brito:2023fwr}
we studied the stability of solutions within this model restricted to spherical dynamics. Although we have found stable solutions in such conditions,
the question remained regarding the stability of solutions with respect to generic perturbations. 
Therefore, we herein perform 3+1 evolutions
of the configurations found to be stable in spherical symmetry
to ascertain if they still remain robust.

\subsection{Fundamental boson stars}

\begin{figure}[tbhp]
    \centering
\includegraphics{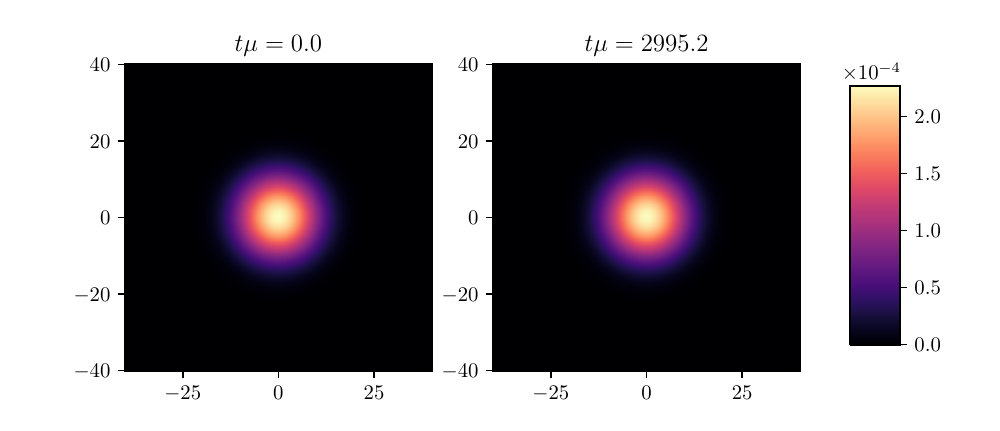}
    \caption{Initial and final state (during the time evolution) of solution II (see~\Cref{solutions}), with $n=0,\ \omega=0.90,\ \Lambda=125$. The snapshots show the $xy$ plane (in units of $\mu$) and the colour code is assigned to the energy density of the scalar field ($\mu^{-2}\rho_K$).}
    \label{ground-stable}
\end{figure}
\begin{figure}[htbp]
    \centering
    \input{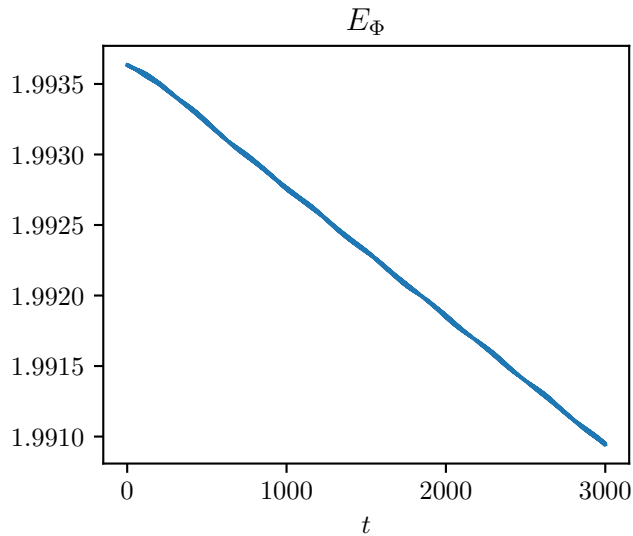}
    \caption{Evolution of the scalar field energy $E_\Phi$ of the stable star shown in \Cref{ground-stable}.}
    \label{komar-mass}
\end{figure}
\begin{figure}[htbp]
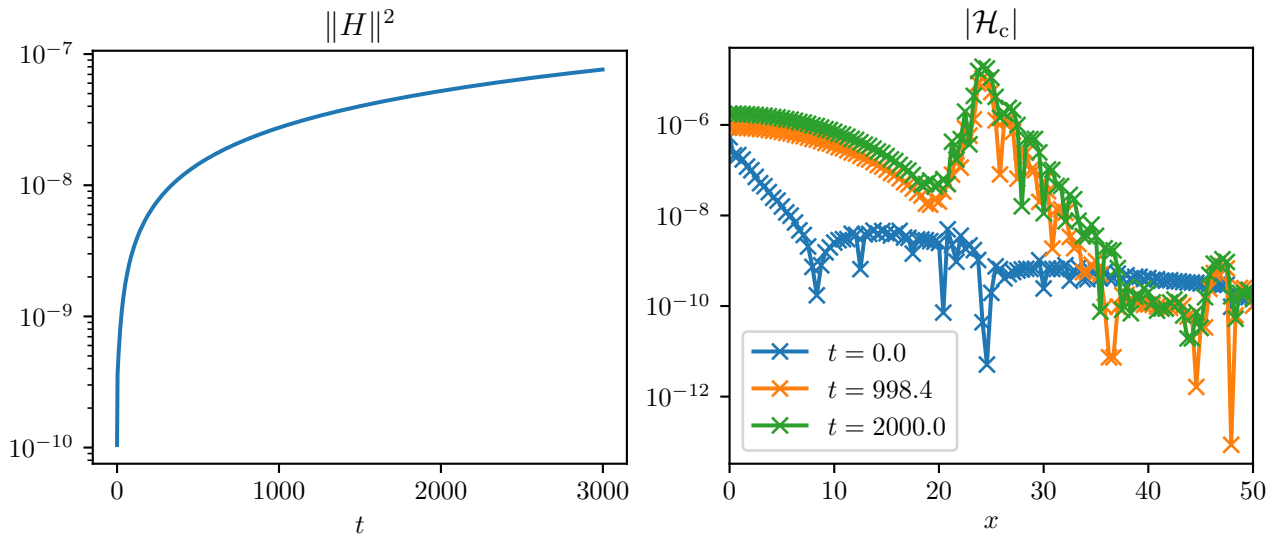

    \centering
    \input{plots/ground_Hamiltonian_L2.pgf}\input{plots/ground_Hamiltonian.pgf}
    \caption{Violation of the L2-norm of the Hamiltonian (left) and of the Hamiltonian
        constraint along $x$ in a constant $t$ hypersurface (right) for the same star shown in Figures~\ref{ground-stable} and \ref{komar-mass}.}
    \label{ground-state-ham}
\end{figure}

It is well known that boson stars in the fundamental-state and in the right frequency range -- from the maximal frequency until the maximum of the ADM mass --  are stable against all kinds of
perturbations, so one can use these solutions to test our code. The evolution of one such star shows that it is stable -- \Cref{ground-stable} -- 
losing only minute amounts of mass (around $0.1\%$) -- \Cref{komar-mass} -- throughout the evolution, which, experience shows, is an unavoidable
numerical effect, 
where the energy $E_\Phi$ of the scalar field is computed as in eq.~(12) from~\cite{Cunha:2017wao}.
For completeness, for this case we also show the violation of the Hamiltonian constraint  in \Cref{ground-state-ham}, given by~\cite{Cunha:2017wao}
\begin{equation}
    \mathcal{H}\coloneqq R-K_{ij}K^{ij}+K^2-16\pi\rho=0,
\end{equation}
where $\rho=n^\mu n^\nu T_{\mu\nu}$.
The violation is small at the initial time, and even though it increases slightly during the
evolution it is still within well acceptable values.

\subsection{Excited boson stars} \label{excite-inst}

It was reported in \cite{Brito:2023fwr} that excited boson stars could be made stable, considering only radial perturbations, if a quartic self-interaction was strong enough. In what follows we will evolve such radially
stable configurations and test if they are indeed stable in a full 3+1D evolution.

\begin{figure}[tbhp]
    \centering
    \includegraphics{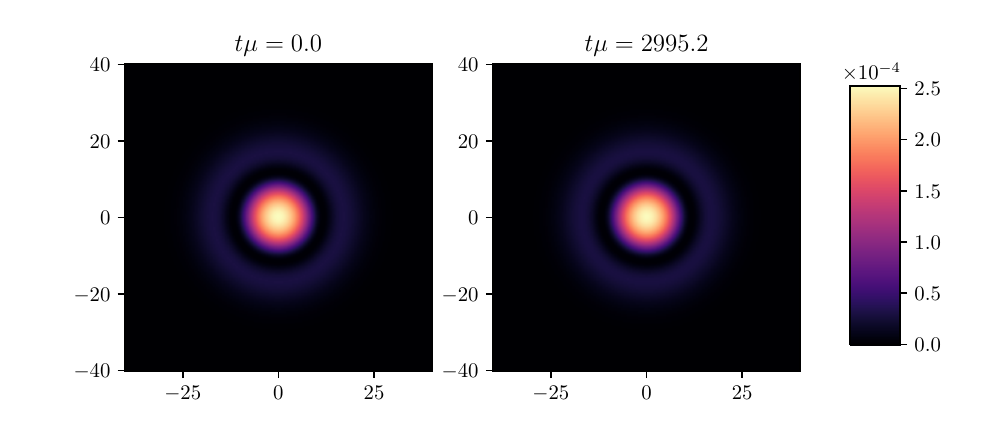}
    \caption{Initial and final state of solution III, with $n=1,\ \omega=0.92,\ \Lambda=100$. Again, the snapshots show the $xy$ plane (in units of $\mu$) and the colour code is assigned to the energy density of the scalar field ($\mu^{-2}\rho_K$).}
    \label{1st-state-stable}
\end{figure}
\begin{figure}[htbp]
    \centering
    \includegraphics[width=0.48\textwidth]{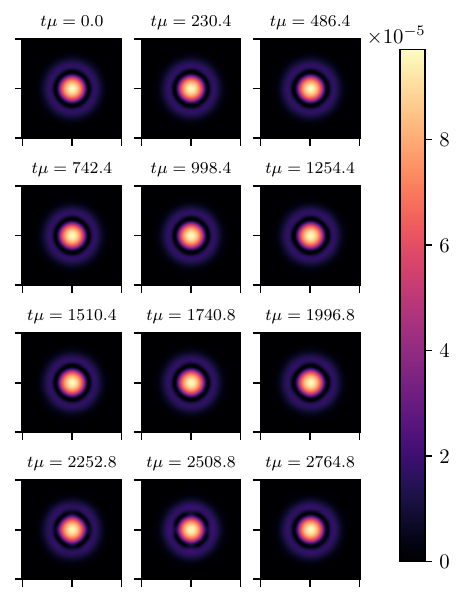}\includegraphics[width=0.51\textwidth]{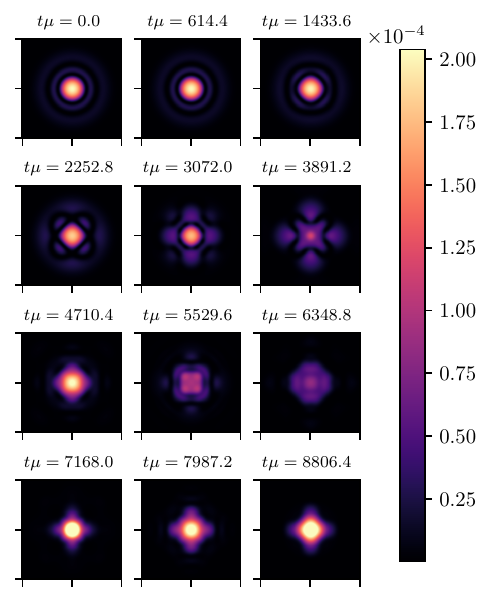}
    \caption{Evolution of solution V, with
$n=1,\ \omega=0.92,\ \Lambda=250$ (left) and solution VI, with $n=2,\ \omega=0.92,\ \Lambda=200$ (right). The snapshots show the $xy$ plane (with grid dimensions $[-100\mu^{-1}\times100\mu^{-1}]$) and the colour code is assigned to the energy density of the scalar field ($\mu^{-2}\rho_K$).}
    \label{n1-w92-l250-evo-n2-w92-l200}
\end{figure}

We first consider star III (see~\Cref{solutions}), with $(n=1,\,\omega=0.92,\,\Lambda=100)$ which was found stable in our previous work on spherical dynamics.
The analysis herein shows that up to $t_{\rm{max}}=3000$ the star shows no visible dynamics -- \Cref{1st-state-stable} -- not even undergoing oscillations for
intermediate times. We suspect, however, that an instability develops for longer evolutions. In fact, this instability (a four-lobe/square pattern aligned with the Cartesian grid) is seen within the same or smaller timescale either increasing $n$ or $\Lambda$, for fixed $\omega$. 
This can be observed in  \Cref{n1-w92-l250-evo-n2-w92-l200}. The left panels show the evolution of a star with larger $\Lambda=250$, solution V. In this case, some non-trivial dynamics can be appreciated (last three snapshots) even though it does not seem to destroy the excited star within the timescale probed by the simulation.
On the other hand, for solution VI, shown in the right panels, a smaller $\Lambda=200$ is seen to be enough to trigger a non-spherical instability which makes the excited star decay into a fundamental, nodeless, state, even though it remains oscillating at the end of the evolution performed. {For $n=1, \Lambda=250$ (left panels), the late snapshots show a faint $m=4$–like deformation and a drift of the radial node, consistent with the growth of a non-radial mode seeded by truncation error and the grid’s discrete symmetries. For $n=2, \Lambda=200$ (right), the deformation is stronger and accompanied by energy-density redistribution between shells—precursors of the chain/ring morphologies seen later in collisions.} Interestingly, spherically symmetric $\ell$-Proca stars~\cite{lazarte2024ell}, which possess nodes in their radial profile, also suffer a similar non-spherical instability~\cite{Lazarte2025}.

The non-spherical instability described above is only present in excited states, but not for the fundamental state, which supports the physical interpretation that these excited states are unstable when considering generic dynamics. This contrasts with the findings in \cite{Sanchis-Gual:2021phr,Brito:2023fwr}
which was, however, limited to spherical dynamics. {The excited configurations were only probed with radial perturbations in such setup; whereas our 3+1D evolutions reveal generic non-spherical instabilities that drive decay to the ground state. Furthermore, in the spherical case higher values of $\Lambda$ stabilized the boson star, whereas in this case stronger self-interactions decrease the instability timescale.} Notice, however, that the timescale of the instability may be sufficient to allow us to perform head-on collisions of these excited states, in case they do not start too far away from each other. This, indeed, will be considered in the next section.

\section{Boson star head-on collisions and merger} \label{collisions}

Excited stars were found above to possess an instability; nonetheless, it is possible to collide these objects as long as the collision timescale is smaller than the instability timescale. We proceed to analyze the outcome of the head-on collision of boson stars with the same frequency $\omega$ and radial node number $n$.
Depending on where they sit on the corresponding solution curve -- \Cref{M-vs-w} -- the remnant may be either a black hole or a bosonic remnant, avoiding collapse. Let us analyze both cases in the following.

\subsection{Black hole remnant}

\subsubsection{Test case: fundamental boson stars}

The case with $n=0$ is useful as code test. Additionally, choosing solution I, with $\omega=0.90$ in the model with $\Lambda=20$,
provides stars with a small effective radius (around $r_{99}\simeq 12.7$), compared to the size of the numerical grid. Thus one can use smaller
grid refinements, thereby speeding up the evolution and saving computational resources.

\begin{figure}[tbhp]
    \centering
    \includegraphics[width=0.75\textwidth]{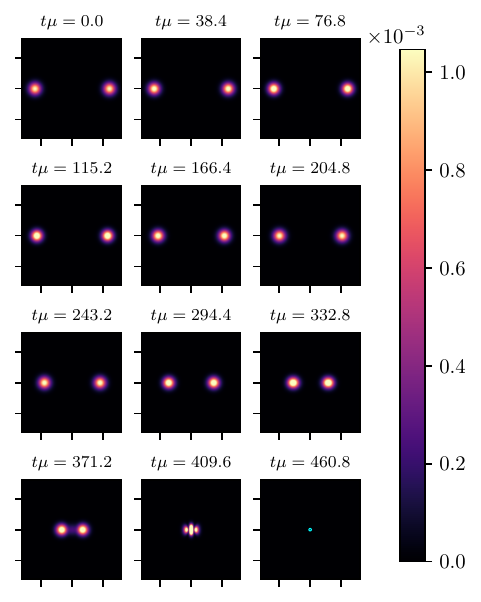}
    \caption{Snapshots in the $xy$ plane ($[-40\mu^{-1}\times40\mu^{-1}]$) of the collision of two fundamental stars (solutions I) with
        $n=0,\ \omega=0.90,\ \Lambda=20$. The colour code is assigned to the Komar energy density of the scalar field ($\mu^{-2}\rho_K$). The
        contour in blue in the final panel is located at the curve $\alpha(t,x^i)=0.2$ to show the approximate position of the apparent horizon.}
    \label{density-test-case}
\end{figure}
To proceed with the head-on collision, two such equal stars are superimposed according to the scheme described above and separated by $r_{s}=60$. The initial ADM mass of the system is
$M_{\mathrm{ADM}}(t=0)\simeq 1.937$ (agreeing with the sum of the isolated mass of both stars).  Then, 
from the numerical simulation, we observe the result of the merger is a black hole -- \Cref{density-test-case} --
with an apparent horizon appearing at $t_{\mathrm{BH}}\simeq 435$, 
with a mass  $M_{\mathrm{BH}}\simeq 1.918$.  The angular momentum of the
black hole is also very close to zero as expected ($a/M_{\mathrm{BH}}=2.02\times10^{-16}$
which is due to numerical errors). 

\begin{figure}[htbp]
    \centering
    \input{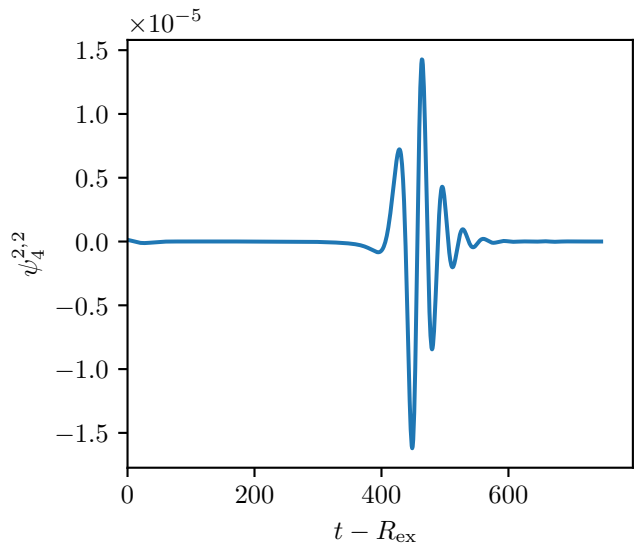}
    \caption{Newman-Penrose scalar $\psi^{2,2}_4$ for a head-on collision of fundamental boson stars with $\omega=0.9$ in a model with $\Lambda=20$.}
    \label{test-case-psi4}
\end{figure}

\begin{figure}[htbp]
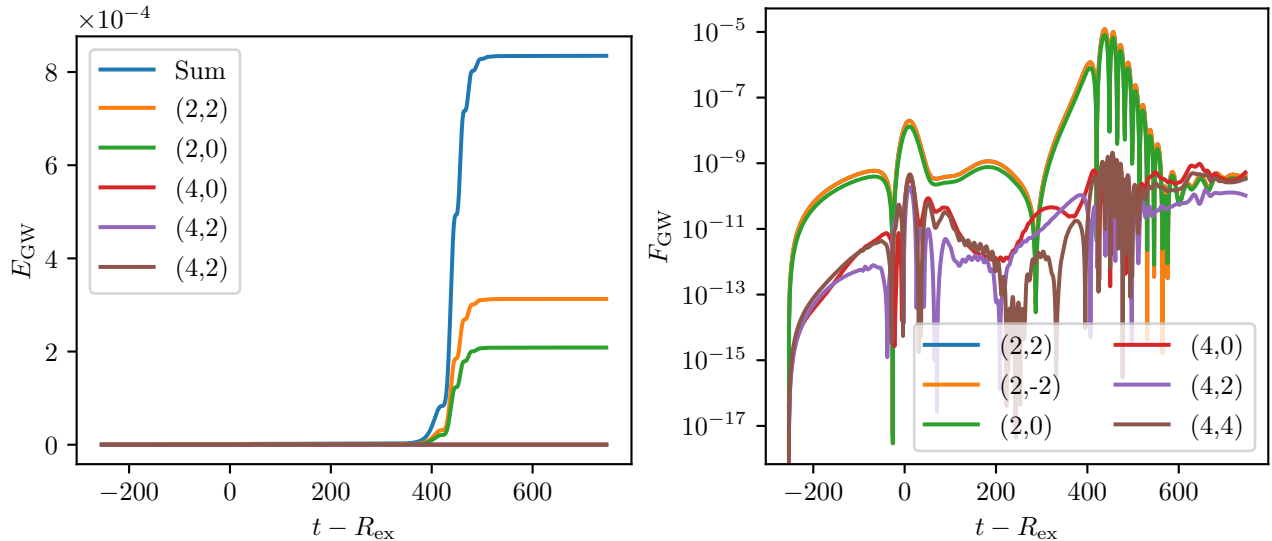

    \centering
    \input{plots/test-energy.pgf}\input{plots/test-power.pgf}
    \caption{Mode and total energy radiated (left panel) and the corresponding power (right panel) in the head-on collision of two fundamental boson stars.}
    \label{test-case-power-and-energy}
\end{figure}

We are able to extract the gravitational wave signal via the Newman-Penrose scalar $\psi_4$, decomposed in harmonics~\cite{Alcubierre-book},
\begin{equation}
\psi_4(x^\alpha)=\sum_{l,m}\psi_4^{l,m}(t,r)Y^{l,m}_{-2}(\theta,\phi), 
\end{equation}
where $Y^{l,m}_{-2}$ are the spin-weighted spherical harmonics. 
\Cref{test-case-psi4} exhibits the quadrupolar mode of the gravitational wave signal, $\psi^{2,2}_4$,  extracted at $R_{\rm{ex}}=254$. The waveform is completely dominated by the merger and ringdown, without noticeable prior power. This is typical in (low energy) head-on collisions forming a black hole.
At any given instant of time we can compute the radiated power by all modes
\begin{equation}
F_\mathrm{GW}\coloneq\frac{\d E_\mathrm{GW}}{\d t}=\lim_{r\to\infty}\frac{r^2}{16\pi}\sum_{l,m}\left|\int_{-\infty}^t\psi^{l,m}_4 \d t'\right|^2.    
\end{equation}
\Cref{test-case-power-and-energy} (left panel) shows the radiated energy for different modes, as well as the total energy. Considering the aforementioned initial ADM mass one can conclude that the total radiated energy is around $E_\mathrm{rad}/M_\mathrm{ADM}\approx 0.04\%$ of the latter. Moreover, after the collision, the radiated energy stalls, as a consequence of the relaxation of the final black hole.  
This conclusion is also corroborated by an analysis of the radiated power -- \Cref{test-case-power-and-energy} (right panel).  We observe the power peaks at the merger and then is quickly suppressed to zero.

\begin{figure}[htbp]
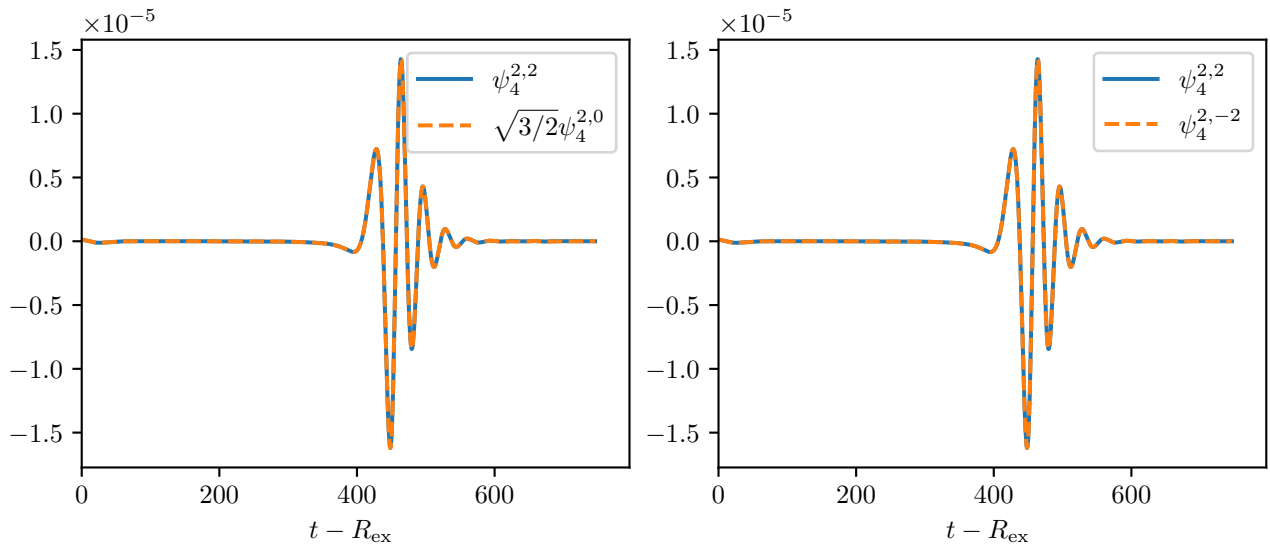

    \centering
    \input{plots/test-gws1.pgf}\input{plots/test-gws2.pgf}
    \caption{Newman-Penrose scalars $\psi^{l,m}_4$ for a head-on collision of fundamental boson stars (solutions I) with $\omega=0.9$ in a model with $\Lambda=20$, verifying \cref{eq-cond1,eq-cond2}.}
    \label{head-on-conds}
\end{figure}

As an additional consistency check of our simulation we
verify whether the conditions \cite{Palenzuela:2006wp,sanchis2019head}
\begin{align}
    \Re(\psi_{4}^{2,2}) & =\Re(\psi_{4}^{2,-2})\label{eq-cond1}            \\
    \Re(\psi_{4}^{2,2}) & =-\sqrt{3/2}\Re(\psi_{4}^{2,0})\label{eq-cond2}
\end{align}
are satisfied. These conditions come from the symmetries of our spacetime and the properties of the spherical harmonics, being valid for any head-on collision of equal objects.
 \Cref{head-on-conds} corroborates that the evolution indeed satisfies \cref{eq-cond1,eq-cond2}.

\begin{figure}[htbp]
    \centering
    \input{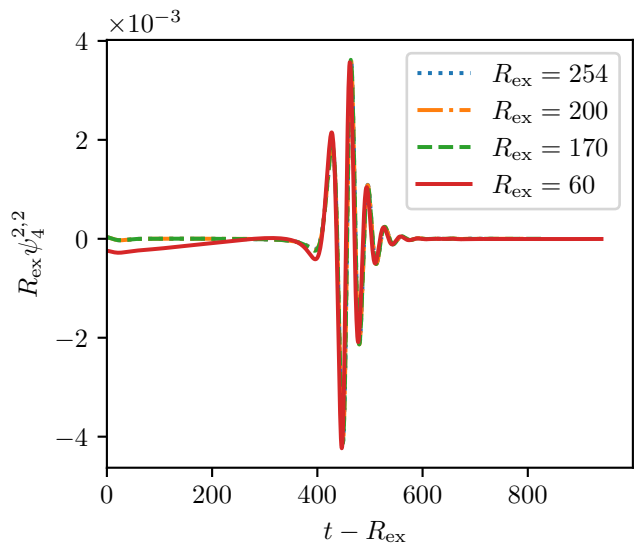}
    \caption{\label{test-case-wavezone} Newman-Penrose scalar scaled by the extraction radius, $R_{\rm ex} \,\psi^{2,2}_4(r)$, for a head-on collision of fundamental boson stars (solutions I) with $\omega=0.9$ in a model with $\Lambda=20$, taken at different extraction radii.}
\end{figure}

We remark that the extraction radius for the analysis in \Cref{test-case-psi4} is 
 already inside the wave-zone, as shown by \Cref{test-case-wavezone}.
A large extraction radius removes practically all junk radiation.

\subsubsection{Comparison with different excitation numbers and different \texorpdfstring{$\Lambda$}{Lambda}}

\begin{figure}[tpbh]
    \centering
\includegraphics[width=0.75\textwidth]{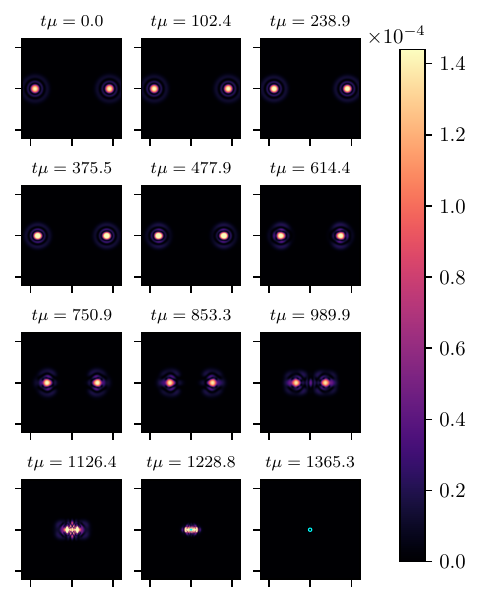}
    \caption{Collision of two stars, solutions VI, with
        $n=2,\ \omega=0.92,\ \Lambda=200$. The snapshots show the $xy$ plane ($[-240\mu^{-1}\times240\mu^{-1}]$) and the colour code is assigned to the Komar energy density of the scalar field ($\mu^{-2}\rho_K$). The
        contour in blue is located at the curve $\alpha(t,x^i)=0.2$ to show the approximate position of the apparent horizon.}
    \label{density-2nd-state}
\end{figure}

\begin{figure}[htbp]
    \centering
    \input{plots/comp_gw_n.pgf}
    \caption{Newman-Penrose scalar $\psi^{2,2}_4$ for two head-on collisions of boson stars with $n=1,2$ (solutions IV and VI) $\omega=0.92$ in a model with $\Lambda=200$, with both waveforms shifted to coincide with their collision times.}
    \label{diff-ex-no-psi4}
\end{figure}
We now compare the collision of boson stars with $\Lambda=200,\omega=0.92$ and $n=1$ (solutions IV)
and $n=2$ (solutions VI). In both evolutions the stars are separated by $4r_{99}$. The snapshots of the evolution for the solutions IV are depicted in \Cref{density-2nd-state}, up to the formation of a black hole.
For the boson stars collisions mentioned above,
the waveforms for each case -- \Cref{diff-ex-no-psi4} --  
are qualitatively
similar, except at the beginning of the merger, where the different morphology of the stars has an impact.
Comparing the $n=1$ and $n=2$  collisions, more initial oscillations are visible for the latter, but the amplitude
of both signals is practically the same ($\simeq 4\times10^{-6}$). Overall, it seems there is no clear way to assign the
features observed in the signal to the parameters of the solution.

\begin{figure}[htbp]
    \centering
    \input{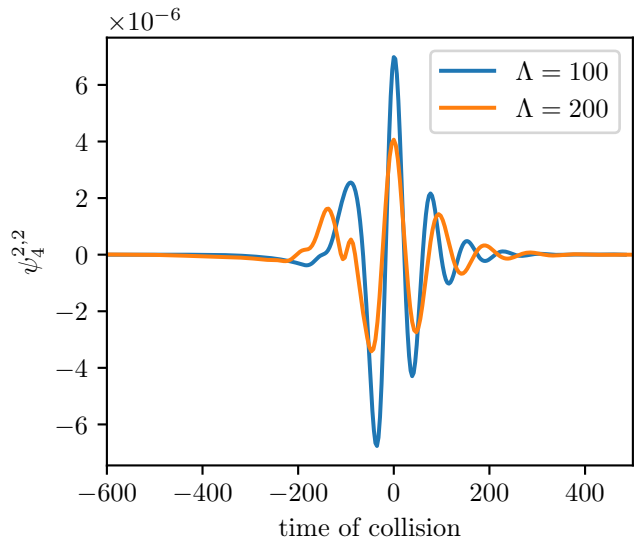}
    \caption{Newman-Penrose scalar $\psi^{2,2}_4$ for two head-on collisions of boson stars with $n=1$, $\omega=0.92$ in a model with $\Lambda=100,200$ (solutions III and IV), with both waveforms shifted to coincide with their collision times.}
    \label{diff-lambda}
\end{figure}

Fixing now the excitation number $n$, one can assess the impact of $\Lambda$. Comparing collisions of solutions III and collisions of solutions IV, with, respectively, $n=1,\omega=0.92$, $\Lambda=100$
and $\Lambda=200$, again with a separation of $4r_{99}$, both waveforms have again a similar pattern -- \Cref{diff-lambda}. 
A noteworthy feature of these collisions is that
the amplitude of the waveform decreases for the evolution with $\Lambda=200$. {This may be attributed to the larger effective radius of the stars with $\Lambda=200$, which causes the merger process to take longer, due to the increased dynamical friction.}

\subsection{Boson star remnant}

Not all boson star collisions give rise to the formation of a black
hole. If the stars are sufficiently non-compact, the resulting object
might have a mass below the maximum ADM mass for boson stars within a model with a certain $\Lambda$, supporting
in principle a new remnant boson star. Again, we shall collide two equal stars, considering different 
excitation numbers, namely $n=0,1,2$. The parameters of the solutions are chosen so that the sum
of the isolated mass of the two stars is smaller than the maximum
allowed mass for boson stars in the respective branch. This forces us to use low compactness
stars. For these stars, the instability found in \Cref{excite-inst} does not appear at the very least for $t_\mathrm{max}=3000$, since these solutions are dilute.

\subsubsection{Collision of fundamental state boson stars}

\begin{figure}[tpbh]
    \centering
    \includegraphics[width=0.75\textwidth]{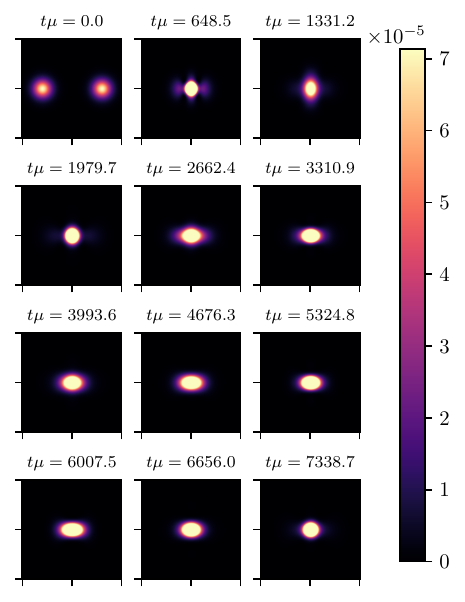}
    \caption{Collision of two stars with
        $n=0,\ \omega=0.98,\ \Lambda=0$ (solutions VII). The snapshots show the $xy$ plane ($[-100\mu^{-1}\times100\mu^{-1}]$) and the colour code is assigned to the Komar energy density of the scalar field ($\mu^{-2}\rho_K$).}
    \label{snapshots-ground-state}
\end{figure}

Consider now the collision of two solutions VII, with $n=0,\omega=0.98,\Lambda=0$. The sum of the isolated
mass of the stars ($M_\mathrm{ADM}(t=0)\simeq 0.67$), which albeit larger than the maximum ADM mass ($\simeq 0.632$) for fundamental mini-boson stars, it is close enough such that the energy loss via gravitational waves and gravitational cooling can bring the remnant's mass below such maximal mass. 
We can track the evolution of the
system in \Cref{snapshots-ground-state}.
We are left with
a highly perturbed nodeless boson star, still undergoing oscillations. At the end of the evolution, we find that the 
final mass of the system is $M_\mathrm{ADM}(t_{\rm{last}}) \simeq 0.53$ and the dominant frequency changes to $\omega=0.94$, placing it right on top of the solution curve for fundamental
boson stars with $\Lambda=0$. 
This is a hint that the final state of this evolution will be a fundamental spherical mini-boson star, even though reaching a static equilibrium may be a very long process.

\subsubsection{Collision of excited stars and boson star chains} \label{chains-rings}

\begin{figure}[tpbh]
    \centering
    \includegraphics[width=0.75\textwidth]{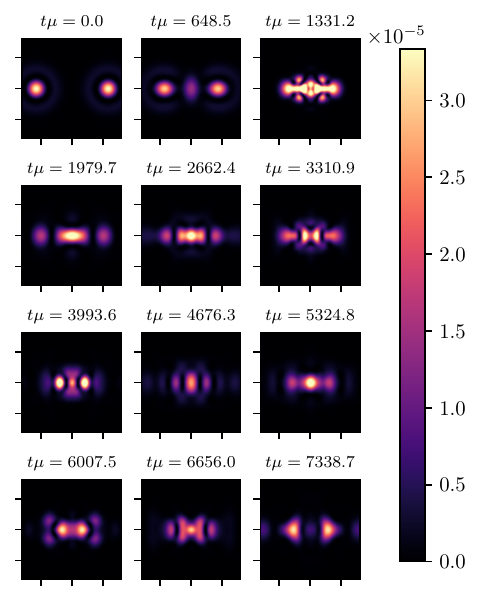}
    \caption{Collision of two stars with
        $n=1,\ \omega=0.98,\ \Lambda=100$ (solutions VIII). The snapshots show the $xy$ plane ($[-160\mu^{-1}\times160\mu^{-1}]$) and the colour code is assigned to the Komar energy density of the scalar field ($\mu^{-2}\rho_K$).}
    \label{snapshots-1st-state}
\end{figure}

\begin{figure}
    \centering
    \includegraphics[width=0.33\textwidth]{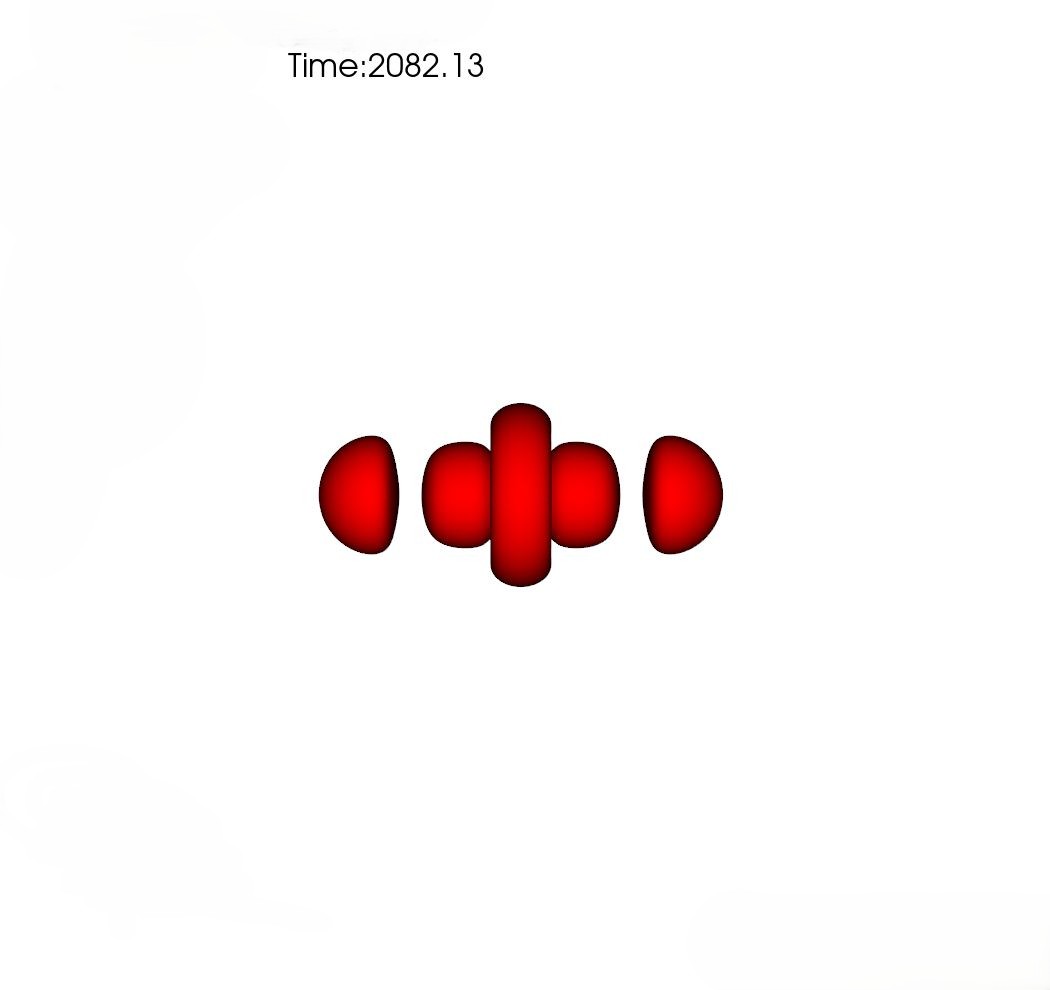}\includegraphics[width=0.33\textwidth]{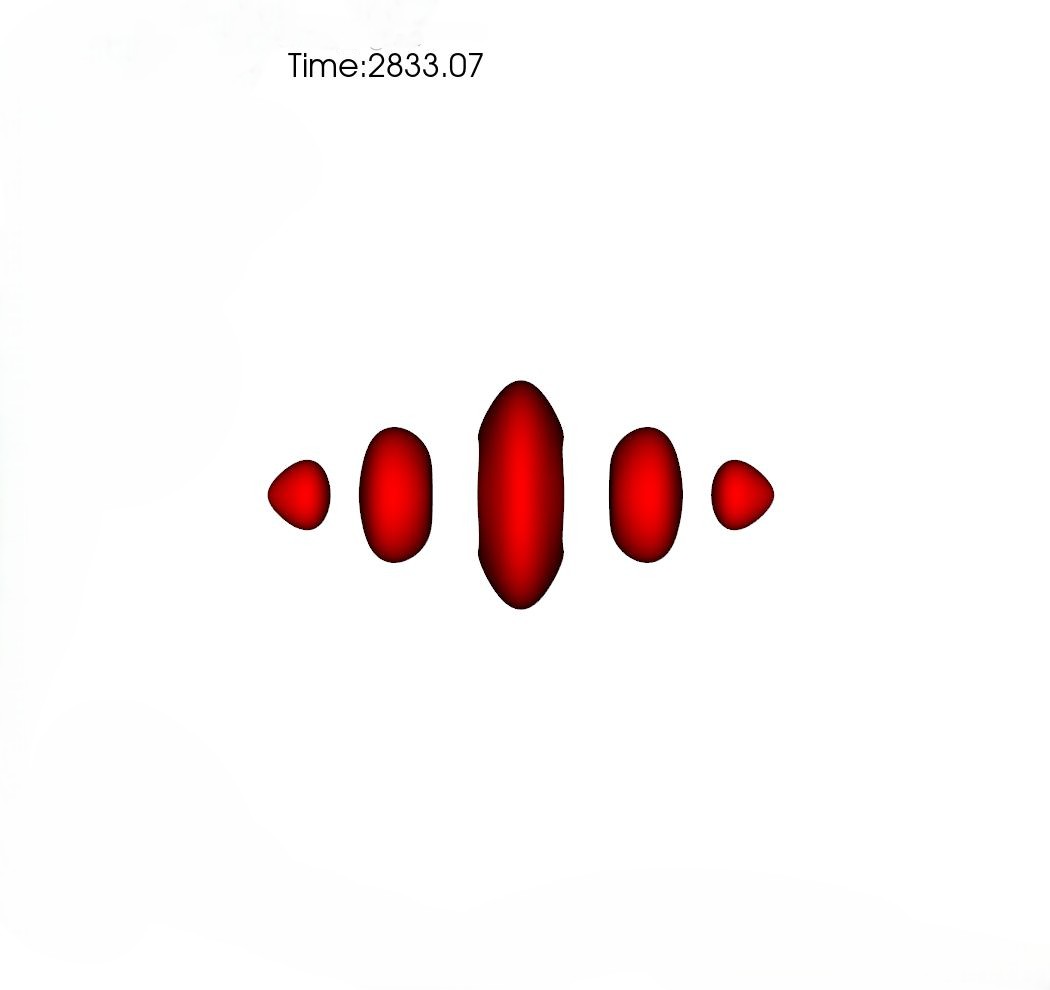}\includegraphics[width=0.33\textwidth]{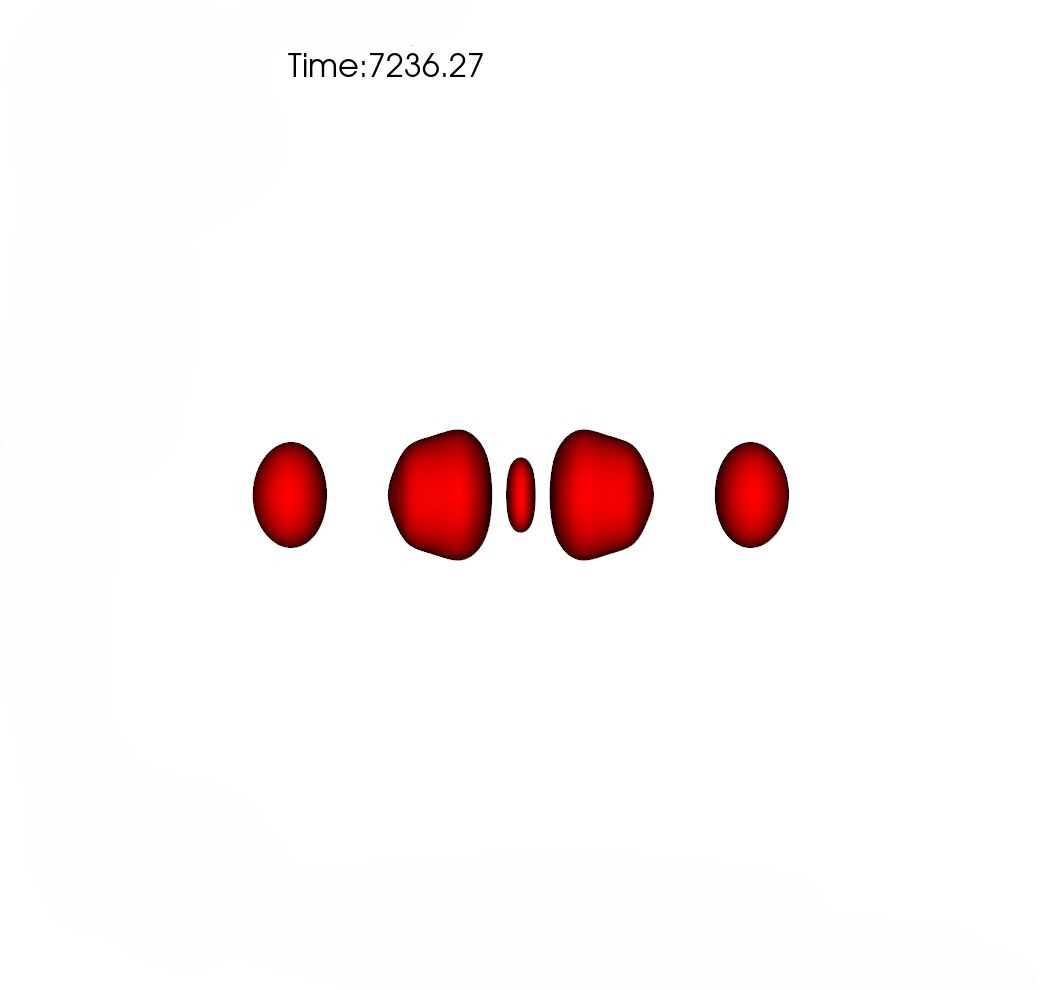}
    \caption{Outer contour surface ($\mu^{-2}\rho_K = 3.03\times10^{-5}$) of the scalar field energy density
        for the head-on collision of two stars with $n=1,\ \omega=0.98,\ \Lambda=100$ (solutions VIII).}
    \label{3d-snapshots-n1-lambda100-chains}
\end{figure}

Consider now the collision of excited boson stars, more specifically the case of solutions VIII, with $n=1,\omega=0.98,\Lambda=100$. Given that the ADM mass is $M_\mathrm{ADM}(t=0)\simeq 1.904$, and that the maximum mass for the solution curve of these configurations  is $M_\mathrm{max} \simeq 2.80$, it is expected that a black hole will not be formed, since the initial mass is already smaller than the maximum mass allowed for this model. Here, the situation is, however, different from the previous case, since the
stars never coalesce into a single object. In fact, the result of the collision is still very dynamical. Remarkably, it appears to evolve sweeping different equilibrium configurations found in~\cite{Liang:2025myf}, describing
boson star chains and rings which, curiously, bifurcate, as equilibrium solutions, from excited spherical boson stars --  \Cref{snapshots-1st-state,3d-snapshots-n1-lambda100-chains}. {These chain, ring-like and mixed configurations bifurcate from spherically symmetric excited boson stars after perturbing such solutions with axisymmetric perturbations with a certain angular momentum and amplitude.}

\begin{figure}[htbp]
    \centering
    \includegraphics[width=0.33\textwidth]{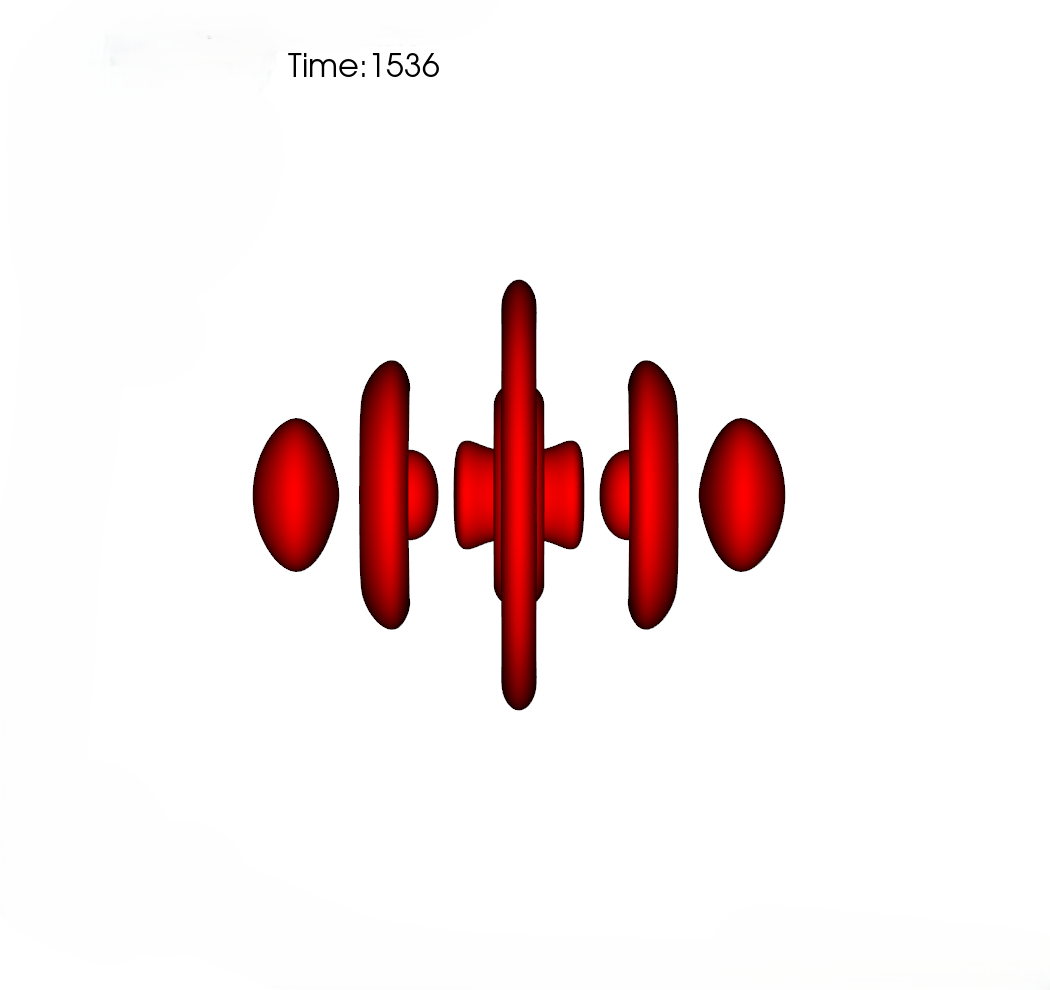}\includegraphics[width=0.33\textwidth]{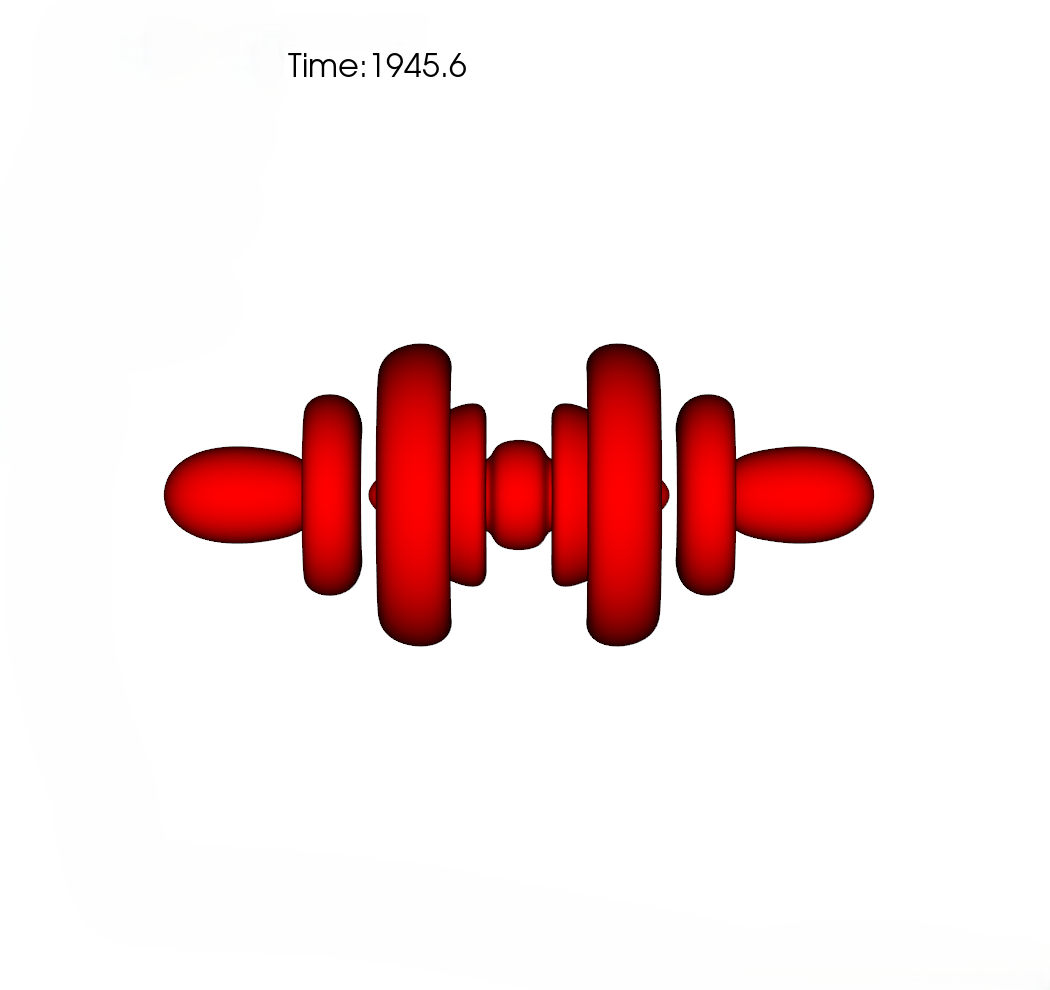}\includegraphics[width=0.33\textwidth]{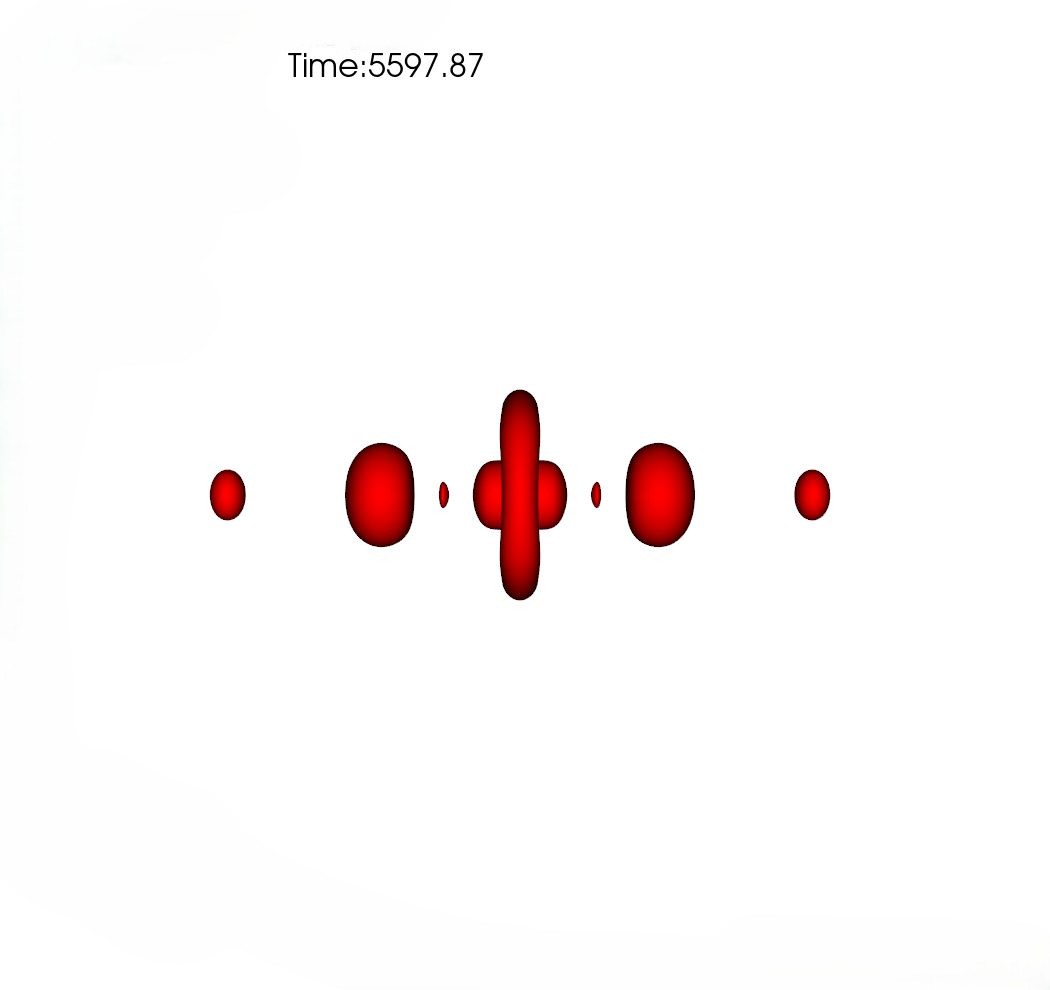}
    \caption{Outer contour surface ($\mu^{-2}\rho_K = 4.98\times10^{-5}$) of the scalar field
        energy density for the head-on collision of two stars with $n=2,\ \omega=0.98,\ \Lambda=100$ (solutions IX).}
    \label{3d-snapshots-n2-lambda100-chains}
\end{figure}

This behaviour can also be observed for other excited  stars with a different $\Lambda$.
Comparing the collision
between stars with $n=1$ and collisions with stars with $n=2$ we find that the latter results in an even more dynamical outcome compared to the
former, allowing us to see the chains and rings more clearly -- \Cref{3d-snapshots-n1-lambda100-chains,3d-snapshots-n2-lambda100-chains}.
The $n=2$ case resembles again a dynamical superposition of chains and rings but with more nodes, as the ones observed in~\cite{Liang:2025myf} bifurcating from the spherical excited boson stars with more nodes as well, sweeping over the solution space as time goes on. {The end state is not reached during  our evolutions, but since the star is radiating gravitational energy, it is possible that it will settle down in a stationary state or disperse completely, unless a truly stable non-spherical equilibrium exists in this model. Our runs are not long enough to differentiate these possibilities conclusively.}
Thus, we find that for excited stars we cannot get a spherical bosonic remnant, even approximately, let alone an excited boson
star.  This is a crucial difference when compared with their fundamental counterparts.

\section{Discussion and conclusion}

In this paper, we showed that scalar spherical excited boson stars, which might be stabilized with respect to radial perturbations by strong enough self-interactions, become unstable when considering generic perturbations, decaying into a fundamental state configuration. 
The timescale of the instability depends on the parameters of the solution, but for generic perturbations, stronger self-interactions trigger the instability sooner, instead of quenching it. This can be compared with the analysis in \cite{Sanchis-Gual:2021phr,Brito:2023fwr}, where stronger self-interactions stabilized these excited boson stars, restricted to spherical dynamics. {This corroborates the findings in \cite{Nambo:2024gvs}, although our evolution timescale and/or the region of the parameter space we have scanned do not allow us to verify the existence of a narrow instability band for the $n=1$ boson stars pointed out in~\cite{Nambo:2024gvs}.}
Yet, the timescale of the non-spherical instability can be large enough compared to that of a head-on collision of these objects, for the chosen solution. This is certainly the case for the very dilute stars used in \Cref{chains-rings}. {This poses problems for the astrophysical viability of these objects as compact objects, due to their non-spherical instabilities, unless the constituent boson is extremely light to allow a very long lifetime.}

Considering the boson star collisions, in cases where the sum of the mass of both stars is greater than the maximum permitted ADM mass,
we obtain a black hole remnant. We obtained the respective gravitational wave signal and concluded
that changing the excitation number and the strength of the self-interaction does change the form of
the signal in a qualitative way. {We are not able, however, to assign the features of the waveforms to the parameters of the solution, other than the impact of the effective radius on the collision due to increased dynamical friction.}

On the other hand, colliding stars such that their combined (final) mass is below the maximum
allowed ADM mass, we  obtained a bosonic remnant avoiding gravitational
collapse.
We found that the collision of fundamental boson stars results in an oscillating spherical boson
star, in fact, migrating to another solution in its solution curve. The situation is different
regarding excited states. The remnant never reaches an equilibrium state, but remains oscillating, momentarily forming chains and rings observed as equilibrium solutions in~\cite{Liang:2025myf}. This is a key difference compared to fundamental boson stars: 
the latter can be formed dynamically in collisions, whereas excited states cannot be
formed in this way. {The only way found so far to dynamically form these excited states (at least for the $n=1$ state) is through dilute Gaussian initial data, although already with a node as found in~\cite{Sanchis-Gual:2021phr}. Yet, this was only performed considering spherical dynamics, so it is possible in a 3+1D evolution such states cannot be formed in this way at all.}

\begin{acknowledgments}
We thank Jordan Nicoules for help and discussions.
This work is supported by CIDMA under the Portuguese Foundation for Science and Technology (FCT, \url{https://ror.org/00snfqn58}) Multi-Annual Financing Program for R\&D Units, grants UID/4106/2025,  UID/PRR/4106/2025”, 2022.04560.PTDC (\url{https://doi.org/10.54499/2022.04560.PTDC}) and 2024.05617.CERN (\url{https://doi.org/10.54499/2024.05617.CERN}). 
M.B.\ is supported by the FCT grant 2022.09704.BD (\url{https://doi.org/10.54499/2022.09704.BD}). N.S.G.\ acknowledges support from the Spanish Ministry of Science and Innovation via the Ram\'on y Cajal programme (grant RYC2022-037424-I), funded by MCIN/AEI/10.13039/501100011033 and by ``ESF Investing in your future'', and by the Spanish Agencia Estatal de Investigaci\'on (Grant PID2021-125485NB-C21) funded by
MCIN/AEI/10.13039/501100011033 and ERDF A way
of making Europe.
M.Z.\ is supported through FCT grant 2022.00721.CEECIND (\url{https://doi.org/10.54499/2022.00721.CEECIND/CP1720/CT0001}).
This work has further been supported by the European Horizon Europe staff exchange (SE)
programme HORIZON-MSCA-2021-SE-01 Grant No. NewFunFiCO-101086251. 
The authors acknowledge the computer resources provided by the Red Espa\~nola de Supercomputaci\'on (Tirant, MareNostrum5, and Storage5) and the technical support from the IT departments of the Universitat de Val\`encia and the Barcelona Supercomputing Center (allocation grants RES-FI-2023-1-0023, RES-FI-2023-2-0002, RES-FI-2024-2-0012, and RES-FI-2024-3-0007, and 2024.07059.CPCA.A3 (\url{https://doi.org/10.54499/2024.07059.CPCA.A3})), as well as
the Navigator Cluster at the LCA in U.~Coimbra and the Deucalion supercomputer at the Minho Advanced Computing Center through projects 2022.15804.CPCA.A2 and 2024.07872.CPCA.A2 (\url{https://doi.org/10.54499/2024.07872.CPCA.A2}).
\end{acknowledgments}

\appendix

\section{Convergence tests}

\subsection{Single boson star initial data and evolution convergence}

\begin{figure}[htbp]
    \centering
    \input{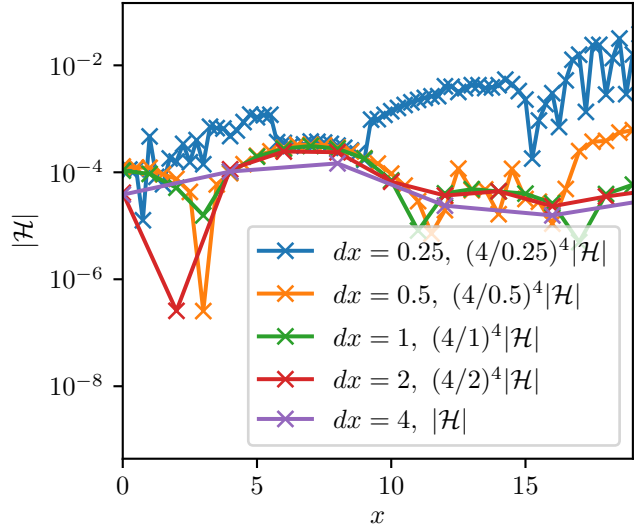}
    \caption{Hamiltonian constraint convergence over the hypersurface $t=0$ for the star with $n=2,\ \omega=0.92,\ \Lambda=200$.}
    \label{single-init-data}
\end{figure}

We test the convergence of the initial data to see that the code is
interpolating the initial data correctly. We do not change the resolution
of the initial data files themselves, but only the resolution of the
evolution grid (without refinement levels). This can be seen in \Cref{single-init-data}.
We obtain a convergence of order 4, as expected, for all resolutions,
except the highest one, perhaps due to increasing numerical truncation
errors {and the fact that we are not changing the resolution of our initial data file}. We obtain the same result for the $yy$ and $zz$ axis which is consistent since our solution
is spherically symmetric.

\begin{figure}[htbp]
    \centering
    \input{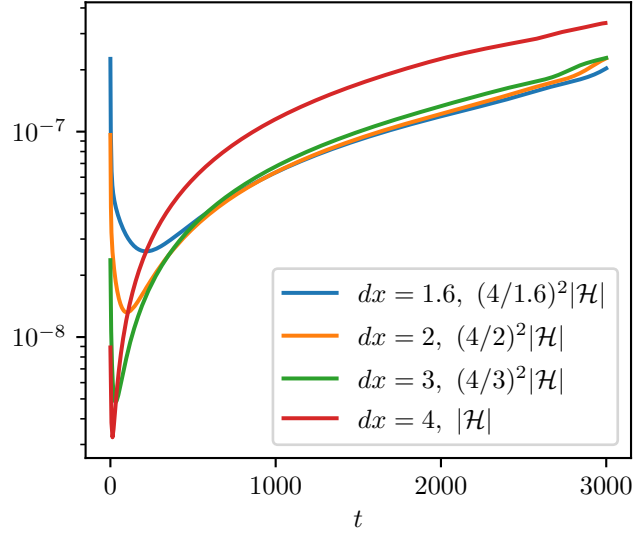}
    \caption{L2-norm of the
        Hamiltonian constraint convergence over time for the star with $n=2,\ \omega=0.92,\ \Lambda=200$.}
    \label{single-init-data-time}
\end{figure}

In order to test convergence during evolution we plot the L2-norm
of the Hamiltonian constraint for four different resolutions, and we
obtain a convergence order of 2 -- \Cref{single-init-data-time}.
We see that we have no convergence for the lowest resolution because it is too low, but
for the remaining three we obtain the expected result.

\subsection{Convergence of the Newman-Penrose scalar $\psi_4^{l,m}$ for head-on collisions}

\begin{figure}[htbp]
    \centering
    \input{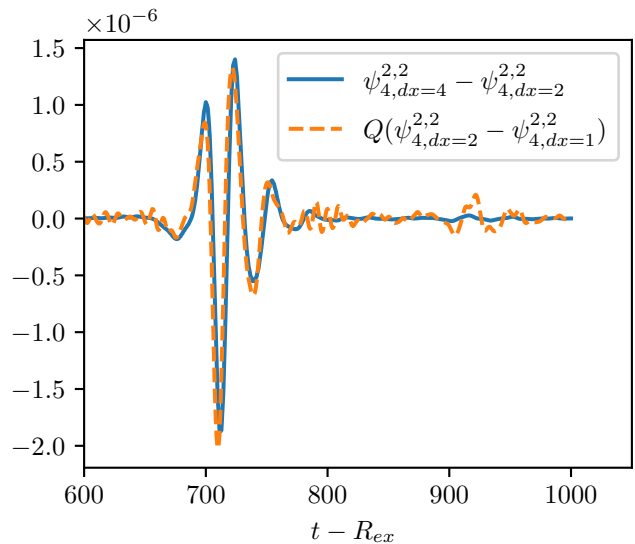}
    \caption{Convergence of $\psi_{4}^{2,2}$ for 3 different resolutions for a
        star with $n=0,\ \omega=0.90,\ \Lambda=20$. A convergence order of about
        $3.5$ was obtained resulting in $Q=11$.}
    \label{conv-gws}
\end{figure}

In order to test the convergence of a collision we cannot use the
Hamiltonian constraint, since in a superposition of the form \cref{sup-eq} the constraint is
not zero because the initial data does not exactly solve the Einstein
equations, so some other function must be used. We choose the wave-mode
$\psi_{4}^{2,2}$ and compare it with three different resolutions,
$\d x=4,2,1$, as in \Cref{conv-gws}.
We obtain a convergence of order around $3.5$, with $Q=11$, where \[Q=\frac{h_1^N-h_2^N}{h_2^N-h_3^N},\] which is consistent with our numerical framework.

\bibliographystyle{aipnum4-2}
\bibliography{refs}

\end{document}